\documentclass[aps,pre,onecolumn,floatfix]{revtex4}

\usepackage{epsf}
\usepackage{subfigure}
\usepackage{amsmath}
\usepackage{amssymb}
\usepackage{bm}
\usepackage{graphicx}
\usepackage{epstopdf}
\newcommand{\be}{\begin{equation}}
\newcommand{\ee}{\end{equation}}
\newcommand{\bea}{\begin{eqnarray}}
\newcommand{\eea}{\end{eqnarray}}

\newcommand{\parent}[1]{\left( #1 \right)}

\begin{document}

\title{\bf What does a large deviation look like?}

\author{David Andrieux}


\begin{abstract}
Large deviation theory quantifies the occurence of events that deviate from the average behavior of a system. 
Such events arise from non-typical trajectories of the dynamics. 
In this note we derive the time evolution of these rare trajectories.
\end{abstract}

\maketitle

\section{Problem fomulation}

A large deviation calculation typically answers the question: What is the probability to observe, say, a value $\xi$ of the entropy production? (or, more accurately, what is the exponential rate $I(\xi)$ at which this probability decreases in time) \\

Here the questions we investigate are the following: {\it What does the dynamics leading to such a deviation $\xi$ look like? Which trajectories are associated to such deviations?} \\

We adress these questions in the context of Markov chains and derive the effective dynamics governing the large fluctuations of size $\xi$.

\subsection{Markov chains}
\label{lumpedmarkovchains}

We consider a Markov chain characterized by a transition matrix $\hat{G} = \parent{G_{ij}} \in \mathbb{R}^{N\times N}$ on the finite state space $\Omega$.
The probability distribution $\pmb{p} = (p_1, p_2, \ldots, p_N)$ evolves in discrete time steps according to
\bea
\pmb{p}(n+1) = \pmb{p}(n) \hat{G} \, .
\eea
The evolution operator $\hat{G}$ is stochastic, i.e. it is non-negative ($\hat{G} \geq 0$) and its rows sum to one ($\sum_{j} G_{ij} = 1)$.
We assume that the Markov chain is primitive, i.e., there exists an $n_0$ such that $G^{n_0}$ has all positive entries. 
This guarantees that $\hat{G}$ has a unique stationary distribution $\pmb{\pi}$ such that
\bea
\pmb{\pi} = \pmb{\pi} \hat{G} \, .
\eea

\subsection{Large deviations of the entropy production}

For concreteness, we consider the large deviations of the entropy production. The entropy production  is a fluctuating quantity measuring the dissipation occurring along a trajectory $i_0 \rightarrow i_1 \rightarrow \ldots \rightarrow  i_k \rightarrow \ldots \rightarrow i_n$ of the system. It takes the form 
\begin{eqnarray}
S(n) \equiv  \ln \parent{ \prod_{k=0}^{n-1}  \frac{G_{i_k i_{k+1}}}{G_{i_{k+1}i_k} } } \, .
\end{eqnarray}
At long times, its probability distribution is characterized by the rate function 
\begin{eqnarray}
I(\xi) = \lim_{n\rightarrow \infty} - \frac{1}{n} \ln {\rm Prob} \Big[ S(n)/n = \xi\Big]  \, .
\label{ldf}
\end{eqnarray}

An equivalent description is given in terms of the generating function 
\begin{eqnarray}
Q(\lambda)  = \lim_{n\rightarrow \infty} - \frac{1}{n} \ln \left \langle e^{-\lambda S(n)} \right \rangle \, .
\label{Q}
\end{eqnarray}
The generating function is related to the largest eigenvalue $\rho(\lambda)$ of the operator 
\begin{equation} 
\parent{ \hat{L}_\lambda }_{ij}=  G^{1-\lambda}_{ij} G^{\lambda}_{ji}
\label{L}
\end{equation}
as follows: 
\bea
Q(\lambda) = - \ln \rho(\lambda) \, .
\eea
The operator (\ref{L}) is non-negative but not stochastic: $\sum_{j} (\hat{L}_\lambda)_{ij}  \neq 1$ when $\lambda \neq 0$.
When $\lambda=0$ we recover the original evolution operator: $\hat{L}_{\lambda = 0} = \hat{G}$.

The descriptions in terms of the rate function $I$ and the generating function $Q$ are related through the Legendre transform 
\bea
Q(\lambda) = \min_\xi [I(\xi) + \lambda \xi] \, .
\label{legendre}
\eea
 These functions quantify the occurence of events of size $\xi$. The goal of this paper is to go a step further and reveal their dynamical properties.

\section{Large deviations are characterized by an extremal principle}

We consider the space of stochastic matrices compatible with an irreducible non-negative matrix $\hat{A} = (A_{ij}) \in \mathbb{R}^{N\times N}$. A stochastic matrix $\hat{P} = (P_{ij}) \in \mathbb{R}^{N\times N}$ is said to be compatible with $\hat{A}$ if $\hat{P}$ satisfies
\begin{subequations}
\bea 
&1)& P_{ij} \geq 0 \quad {\rm if} \quad A_{ij} \geq 0 \\
&2)& P_{ij} = 0 \quad {\rm if} \quad A_{ij} = 0 \, .
\eea
\end{subequations}

We then have the following\\ 

{\bf Theorem.} 
{\it Let $\Sigma_{\hat{G}}$ be the space of stochastic matrices compatible with $\hat{G}$. Then 
\bea
Q(\lambda) = \min_{\Sigma_{\hat{G}}} \left[ \sum_{i, j} q_i P_{ij} \ln \parent{\frac{P_{ij}}{G_{ij}}} + \lambda  \sum_{i, j} q_i P_{ij} \ln \parent{\frac{G_{ij}}{G_{ji}}} \right] \, .
\label{extr}
\eea
Here $\pmb{q}$ is the stationary probability distribution of $\hat{P}$ ($\pmb{q} = \pmb{q} \hat{P}$).
}\\
 
DEMONSTRATION. 
Direct application of the Theorem 1.2 of Ref. \cite{S11} to the operator $\hat{L}_\lambda$. $\Box$\\
 
Furthermore, the optimal dynamics is given by the following\\

{\bf Theorem.} {\it The equality in (\ref{extr}) holds for the stochastic dynamics
\bea
\hat{P}^*_\lambda = \frac{1}{\rho(\lambda)} {\rm diag} (\pmb{x}_\lambda)^{-1}\ \hat{L}_\lambda \ {\rm diag}(\pmb{x}_\lambda)
\label{Popt}
\eea 
and its stationary probability distribution
\bea
\pmb{q}^*_\lambda = \frac{\pmb{y}_\lambda \circ \pmb{x}_\lambda}{\pmb{y}_\lambda^{{\rm T}} \pmb{x}_\lambda} \, .
\label{z}
\eea
Here $\pmb{x}_\lambda > 0$ and $\pmb{y}_\lambda > 0$ are, respectively, the right and left eigenvectors of $\hat{L}_\lambda$ corresponding to the
largest eigenvalue $\rho(\lambda)$. ${\rm diag}(\pmb{x})$ denotes the diagonal matrix with $\pmb{x}$ on its diagonal and $\pmb{y} \circ \pmb{x}$ denotes the
vector $(y_1 x_1, . . . , y_N x_N)$.}\\

DEMONSTRATION: Insert (\ref{Popt}) and (\ref{z}) into (\ref{extr}) and check that we have equality. $\Box$\\

\newpage

\section{Interpretation}

The result (\ref{extr}) has a clear physical interpretation. The term
\bea
D[\hat{P}] = \sum_{i, j} q_i P_{ij} \ln \parent{\frac{G_{ij}}{G_{ji}}}
\label{t1}
\eea
 is the entropy production observed when fluctuations generate the statistics $\hat{P}$.  
The term
\bea
J[\hat{P}] = \sum_{i, j} q_i P_{ij} \ln \parent{\frac{P_{ij}}{G_{ij}}}
\label{t2}
\eea
is the Kullback-Leibler distance between the dynamics $\hat{P}$ and $\hat{G}$. It corresponds to the probability rate to observe fluctuations with statistics $\hat{P}$ in the original dynamics.
The generating function at value $\lambda$ thus selects out fluctuations that minimize the deviations from the original dynamics along with the $\lambda$-weighted entropy production.

The physical meaning of the terms (\ref{t1}) and (\ref{t2}) allows the connection with the rate function (\ref{ldf}). 
The optimal dynamics $\hat{P}^*_\lambda$ generates a dissipation $\xi (\lambda) = D[\hat{P}^*_\lambda]$.
Inverting this relation, the rate function is given by 
\bea
I(\xi) = J[\hat{P}^*_{\lambda(\xi)}] \, .
\eea

This establishes the link between the optimal Markov chains and the large deviations. 
The typical trajectories of the optimal chain $\hat{P}^*_\lambda$ 1) lead to an entropy production $\xi (\lambda)$ and 2) appear at the rate $I(\xi)$ in the original dynamics. 
We thus conclude that the large fluctuations of size $\xi(\lambda)$, when observed, appear through the dynamics $P^*_{\lambda(\xi)}$.
This answers our original question on what large devations look like.

\section{Time reversal symmetry and large deviation properties}
\label{timereversal}

The symmetry under time reversal implies a \\

{\bf Fluctuation Theorem.} 
{\it The generating function has the symmetry
\bea
Q(\lambda) = Q(1-\lambda) \, .
\eea
}

DEMONSTRATION: Direct consequence of the relation $\hat{L}_\lambda = \hat{L}^{{\rm T}}_{1-\lambda}$. $\Box$\\

 
How does this symmetry affect the optimal chains? We start by looking at particular cases:\\
\begin{itemize}
   \item $\lambda = 0$: The optimal dynamics is identical to the original chain, $\hat{P}^*_{\lambda=0} = \hat{G}$, and the generating function $Q(\lambda=0)=0$.  
 
 \item $\lambda = 1/2$: The optimal dynamics $P^*_{\lambda=1/2}$ satisfies the detailed balance conditions, leading to $D = \xi = 0$. The generating function thus takes the value $Q(1/2)= I(\xi=0)$. That is, the optimal dynamics minimizes the Kullback-Leibler distance with respect to the original chain, given the detailed balance constraints.

\item $\lambda = 1$: The optimal dynamics is the time reversal of the original chain, $\hat{P}^*_{\lambda=1} = \hat{G}^{{\rm R}}$. The time reversal of a Markov chain $\hat{G}$ is defined as 
\bea 
\hat{G}^{{\rm R}} = {\rm diag}(\pmb{\pi})^{-1} \ \hat{G}^{{\rm T}} \ {\rm diag}(\pmb{\pi})  
\eea
where $\pmb{\pi} > 0$ is the stationary distribution of the chain $\hat{G}$.
\end{itemize}

More generally, we have the following\\
 
{\bf Theorem.} {\it The chains $\hat{P}^*_\lambda$ and $\hat{P}^*_{1-\lambda}$ are time reversal of each other.}\\
 
DEMONSTRATION: Left as an exercise. \\ 
 
This theorem reveals that the paths leading to a given large deviation and those leading to its opposite one are time reversal of each other.

\section{Example: Large deviations of a three-state system}

\begin{figure}[h]
\centerline{\includegraphics[width=10cm]{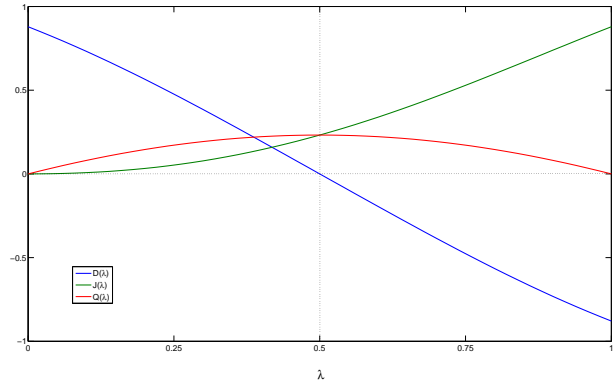}} 
\caption{{\bf Generating function and its decomposition in a three-state system.} The generating function is determined by the rate $J$ and the effective entropy production $D$ of the optimal dynamics $P^*_\lambda$.  The transition probabilities take the values $P_{12}=0.8, P_{13} = 0.2, P_{21}=0.1, P_{23}=0.9, P_{31}=0.7 , P_{32}=0.3$, and $0$ otherwise.}
\label{fig1}
\end{figure}

The generating function is given by $Q(\lambda) = J(\lambda) + \lambda D(\lambda)$, where $D(\lambda) = D[\hat{P}^*_\lambda] =  \xi (\lambda)$ and $J(\lambda) = J[\hat{P}^*_\lambda]$ (Fig. \ref{fig1}). It satisfies the fluctuation theorem (section \ref{timereversal}). 
The effective entropy production $D(\lambda)$ is antisymmetric with respect to $\lambda = 1/2$, where it vanishes due to the equilibrium dynamics (section \ref{timereversal}). At $\lambda = 0$ it takes the value of the average entropy production of the original dynamics $\hat{G}$.
 The rate function $I(\xi)$ is obtained from $J\parent{\lambda(\xi)}$ (not shown).

\begin{figure}[b]
\centerline{\includegraphics[width=20cm]{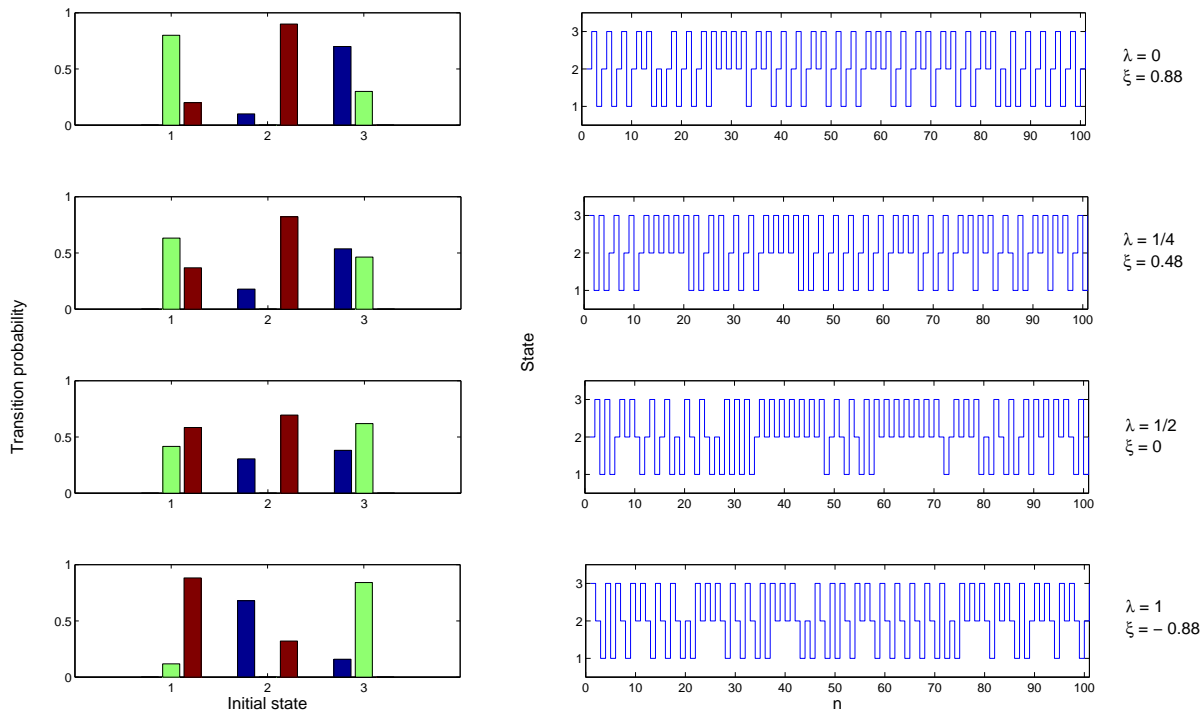}} 
\caption{{\bf Large deviation dynamics in a three-state system}. Each row corresponds to a large deviation of given size (fixed $\lambda$ or $\xi$). The left-hand side displays the transition matrix of the optimal chains [final state = 1 (blue), 2 (green), 3 (red)]. The right-hand side shows the corresponding stochastic trajectories. The parameters take the same values as in Fig. \ref{fig1}.}
\label{fig2}
\end{figure}

Figure \ref{fig2} illustrates the dynamics leading to different rare events, as measured by different values of $\lambda$ or, equivalently, of the entropy production $\xi(\lambda)$. 
The first row corresponds to the original dynamics $\hat{G}$. There is a clear temporal structure in the trajectory, with an average positive circulation (i.e., a tendency to go in the direction $1 \rightarrow 2 \rightarrow 3 \rightarrow 1 \rightarrow \ldots$).
This average tendency disappears progressively until we reach $\lambda = 1/2$, which corresponds to an equilibrium dynamics and fluctuations of vanishing entropy production.
For $\lambda > 1/2$, the fluctuations are, on average, going in the opposite direction (i.e.,  in the direction $1 \rightarrow 3 \rightarrow 2 \rightarrow 1 \rightarrow \ldots$). Finally, the dynamics at $\lambda = 1$ is the time reversal of $\hat{G}$. More generally, large deviations of opposite values are generated through a time-reversed dynamics (section \ref{timereversal}).

\newpage

\section{Conclusions}
\label{Conclusions}

Large deviations can be described in terms of an extremal principle over the space of Markov chains. 
The associated optimal chains describe the dynamical behavior of these large deviations. 
This approach presents neat mathematical properties and offers new insights into the dynamical origin of large deviations. 
The present results readily translate to other quantitiers of interest such as the thermodynamic currents. 
I believe that the knowledge of the paths leading to rare events will play an important role in the description and understanding of nonequilibrium systems.

\vskip 0.5 cm

{\bf Disclaimer.} This paper is not intended for journal publication. 
Therefore, it lacks the delusional claims typically found in published papers about its (simultaneous) relevance to drug design, nanoelectronics, photonics, spintronics or any other fashionable application. 
Also, everyone can sleep at peace without worrying about missed citations and their h-index (and without e-mailing me to complain about them).

%
%
%


\end{document}